\documentclass[prl,aps,twocolumn,amsmath,amssymb,superscriptaddress]{revtex4-1}
\usepackage{graphicx}
\usepackage{dcolumn}
\usepackage{bm}
\usepackage{color}	
\usepackage{pdfsync}
\usepackage{subfigure}

\newcommand{\ket}[1]{\lvert{#1}\rangle}

\begin{document}

\title{Implementation of a Quantum Metamaterial}
\author{Pascal Macha}
\email{Electronic address: p.macha@uq.edu.au}
\altaffiliation{Present address: ARC Centre for Engineered Quantum Systems, University of Queensland, Brisbane 4072, Australia}
\affiliation{Institute of Photonic Technology, P.O. Box 100239, D-07702 Jena, Germany}
\affiliation{Physikalisches Institut, Karlsruhe Institute of Technology, D-76128 Karlsruhe, Germany}

\author{Gregor Oelsner}
\affiliation{Institute of Photonic Technology, P.O. Box 100239, D-07702 Jena, Germany}
\author{Jan-Michael Reiner}
\affiliation{Institut f\"ur Theoretische Festk\"orperphysik and DFG-Center for Functional Nanostructures (CFN), Karlsruhe Institute of Technology, D-76128 Karlsruhe, Germany}
\author{Michael Marthaler}
\affiliation{Institut f\"ur Theoretische Festk\"orperphysik and DFG-Center for Functional Nanostructures (CFN), Karlsruhe Institute of Technology, D-76128 Karlsruhe, Germany}
\author{Stephan Andr\'e}
\affiliation{Institut f\"ur Theoretische Festk\"orperphysik and DFG-Center for Functional Nanostructures (CFN), Karlsruhe Institute of Technology, D-76128 Karlsruhe, Germany}
\author{Gerd Sch\"on}
\affiliation{Institut f\"ur Theoretische Festk\"orperphysik and DFG-Center for Functional Nanostructures (CFN), Karlsruhe Institute of Technology, D-76128 Karlsruhe, Germany}
\author{Uwe H\"ubner}
\affiliation{Institute of Photonic Technology, P.O. Box 100239, D-07702 Jena, Germany}
\author{Hans-Georg Meyer}
\affiliation{Institute of Photonic Technology, P.O. Box 100239, D-07702 Jena, Germany}
\author{Evgeni Il'ichev}
\affiliation{Institute of Photonic Technology, P.O. Box 100239, D-07702 Jena, Germany}
\affiliation{Russian Quantum Center, 100 Novaya St., Skolkovo, Moscow region, 143025, Russia}
\author{Alexey~V.~Ustinov}
\email{Electronic address: ustinov@kit.edu}
\affiliation{Physikalisches Institut, Karlsruhe Institute of Technology, D-76128 Karlsruhe, Germany}
\affiliation{Russian Quantum Center, 100 Novaya St., Skolkovo, Moscow region, 143025, Russia}
\affiliation{National University of Science and Technology MISIS, Leninsky prosp. 4, Moscow, 119049, Russia}

\maketitle
{\bf Manipulating the propagation of electromagnetic waves through sub-wavelength sized artificial structures is the core function of metamaterials 
\cite{Zheludev2010,Soukoulis-Wegener-2011,Zheludev2012}. Resonant structures, such as split ring resonators, play the role of artificial "atoms" and shape the  magnetic response. Superconducting metamaterials moved into the spotlight for their very low ohmic losses and the possibility to tune their resonance frequency by exploiting the Josephson inductance \cite{Ricci2005, Lazarides2007, Anlage2011, Jung2013}. Moreover, the nonlinear nature of the Josephson inductance enables the fabrication of truly artificial atoms \cite{Wilhelm2008,Astafiev2010,You2010}. Arrays of such superconducting quantum two-level systems (qubits) can be used for the implementation of a quantum metamaterial \cite{Rakhmanov2008}.
Here, we perform an experiment in which 20 superconducting flux qubits are embedded into a single microwave resonator. The phase of the signal transmitted through the resonator reveals the collective resonant coupling of up to 8 qubits.
Quantum circuits of many artificial atoms based on this proof-of-principle experiment offer a wide range of prospects, from detecting single microwave photons \cite{Romero2009} to phase switching \cite{Tian2010}, quantum birefringence \cite{Zagoskin2012} and superradiant phase transitions \cite{Lambert2009}.}

The key issue for the implementation of quantum metamaterials based on superconducting qubits is to demonstrate the existence of collective quantum modes corresponding to coherent oscillations of qubits. However, while natural atoms are identical, superconducting qubits are never exactly the same. For instance, the three-junction flux qubits \cite{Mooij1999} used in this work, have a minimal energy level spacing $\Delta$, which is exponentially sensitive to the design parameters 
\cite{Orlando1999}. This makes the fabrication of qubits with similar specifications a challenging problem. Moreover, in a linear qubit chain, which relies on the nearest-neighbour interaction, single off-resonant qubits act as defects and may destroy coherent modes. These drawbacks can be circumvented by using a common cavity for coupling the qubits one by one to a collective cavity mode. 

Exploring the above ideas, we designed and fabricated a sample featuring 20 superconducting aluminium flux qubits embedded into a niobium microwave $\lambda/2 $ cavity (see Fig. 1a). For the used fabrication process a spread of less than 20 \% in the energy gap $\Delta$ and of 5 \% in the persistent current $I$ is achieved for flux qubits \cite{Jerger2011}. The qubit-qubit nearest-neighbour coupling is designed to be negligibly small and the coupling of each qubit to the resonator is chosen small enough, so that only collective effects are expected to be visible. On resonance, when the level spacings of the qubits are equal to that of the resonator, the degeneracy between their states is lifted. This can be monitored by measuring the amplitude and phase of the microwaves transmitted at the resonator frequency. In the case of $n$ mutually non-interacting qubits, an enhancement of the collective coupling by a factor of $\sqrt{n}$ compared to the single-qubit case is expected  \cite{Dicke1954} and has been previously observed for 3 superconducting qubits \cite{Fink2010}. However, the scalability of this effect for a larger number of qubits forming a macroscopic quantum system has not been tested.

The energy splitting between the ground and the excited state of individual qubits can be controlled by changing the external magnetic field. The frequency spacing between the lowest two energy levels for the $i$-th qubit is given by $E_i = \sqrt{\Delta_i^2+\epsilon_i^2}$
\cite{Mooij1999}. By detuning the external magnetic flux $\Phi$ from the degeneracy point of half a flux quantum $\Phi_0/2$ an energy bias $\epsilon_i = 2 I_{i} \cdot \frac{\Phi - \Phi_0/2}{\hbar}$ is provided.
Flux qubits are extremely anharmonic and therefore the influence of their higher energy levels can be safely ignored.

\begin{figure}
\includegraphics[]{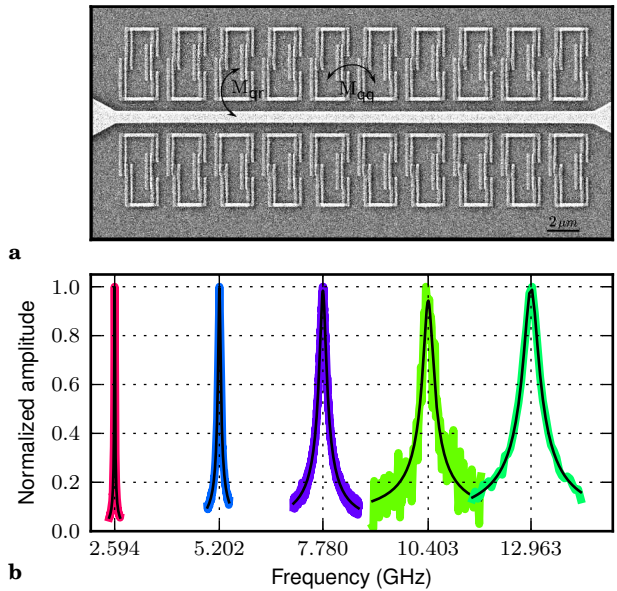}
\caption{\label{fig:SampleHarmonics} {\bf  Micrograph of the sample and spectrum of the resonator. a,} Scanning electron micrograph of the sample showing the central part of the coplanar wave guide resonator. 20 qubit rings are situated between the central conductor line and the ground plane of the resonator on a length of about 20 $\mu$m. The artificial atoms are much smaller than the wavelength of the transmitted signals, which is of the order of the length of the resonator (23 mm). Each qubit is individually coupled to the resonator by the mutual inductance $M_{qr}$ and to its neighbour by $M_{qq}$. 
Even though $M_{qr}$ and $M_{qq}$ are of the same order of magnitude, the effect of direct coupling between the qubits is inhibited by their strong dephasing (see Supplementary Information A). Therefore, the system effectively constitutes $n$ mutually non-interacting spins coupled to the photon field of the resonator. {\bf b,} The transmission amplitude of the resonator at the fundamental mode frequency and the first four harmonics of the resonator. The black lines are fits to Lorentzians. The linewidths $\kappa / (2\pi)$ are 55.5, 216, 715, 950, and 1400 kHz. $\kappa$ is the photon loss rate of the resonator. The widths of the curves are scaled over the frequency axis to a factor of 250 to ensure visibility over this large frequency range. The relative linewidth is to scale.}
\end{figure}
The cavity is formed as a coplanar wave guide resonator (CPWR) with the fundamental mode at $\omega_{1}/2\pi = 2.594$ GHz. The CPWR has higher harmonics $\omega_j \approx j \cdot \omega_1$ with $j=1, 2, 3, ...$, which are accessible through our measurement setup up to $j = 5$ (see Fig. \ref{fig:SampleHarmonics}b). This feature allows for the investigation of resonant interactions at different frequencies. 
We performed the measurements in a dilution refrigerator with a nominal base temperature below 20 mK, resulting in a thermal population of the fundamental mode of about $0.002$ photons.
The phase of the transmission through the sample at the harmonics of the resonator was recorded with a network analyser. A sufficiently small amplitude of the probe signal guaranteed that the average number of photons in the resonator was below unity.
The photon field in the resonator is described by the creation and annihilation operators $a^\dag$ and $a$. The $i$-th qubit Hamiltonian in the energy basis $\{\ket{g_i},\ket{e_i}\}$ can be expressed as $H_q = \frac{\hbar E_i}{2} \sigma_z^i$,  where $\sigma_k^i$ are the Pauli matrices. 
The system of a single resonator mode coupled to $n$ qubits is modelled by the Tavis-Cummings Hamiltonian\;\cite{Tavis1968}
$H = \hbar \omega_j a^{\dag}a + \sum_{i=1}^{n} \left(\frac{\hbar E_i}{2} \sigma_z^{i}+ \hbar g_{\epsilon, ij}(\sigma_+^{i} a + \sigma_-^{i} a^\dag)\right)$
, where $g_{\epsilon, ij} = \left(\Delta_{i}/ E_{i}\right) g_{ij}$ is the transversal coupling of one qubit to the resonator. The bare coupling $g_{ij}$ between qubit $i$ and the resonator mode $j$ can be calculated from the sample's geometry. Numerical calculations for the used geometries revealed a mutual inductance $M_{qr} = (0.5 \pm 0.02)$ pH between a single qubit and the resonator and an inductance $L_r = (11\pm 0.4)$ nH of the resonator. 
Subsequently, the coupling constant follows as $g_{ij} = M_{qr} I_{i} \sqrt{ \omega_j / \hbar L_r}$. The dense packing of the flux qubits prevents inhomogeneous coupling, which would be expected for larger types of superconducting qubits \cite{Hou2012}.
For large dephasing and resonant driving of the oscillator we can use a semiclassical model for the description of the photon field,
\begin{equation}
\dot{ \langle a \rangle } = \left(-\frac{\kappa_j}{2}  +
   \sum_{i=1}^{n} \frac{{g_{\epsilon,ij}}^2 }{\Gamma_{\varphi} - \mathrm{i}\delta_{ij}}   \right) \langle a \rangle -
   \mathrm{i}\frac{f}{2} \;. \label{eq:one}
\end{equation}
$\Gamma_{\varphi}$ is the dephasing rate of the qubits, $f$ the driving strength and $\delta_{ij} = E_i - \omega_j$ is the detuning. We assume that the dephasing rate is the same for all qubits and since the driving is weak we neglected terms of the order of $|\langle a \rangle|^2$ (see Supplementary Information A). In the experiment the phase $\phi$ of the transmission coefficient $t$ of the resonator is monitored, $t \propto \langle a \rangle= |\langle a \rangle| e^{i \phi}$.
The parameters of the qubit-resonator system are in the weak-coupling limit, where $\kappa_j \sim g_{\epsilon,ij}$  and $\Gamma_{\varphi} \gg g_{\epsilon,ij}$. Hence, single qubit anticrossings are not resolvable from the base noise level. 
For $n$ resonant qubits an intermediate regime may be reached, when $\Gamma_\varphi > \sqrt{n} g_{\epsilon,ij}$ and $\kappa < \sqrt{n} g_{\epsilon,ij} $. However, the vacuum Rabi splitting of a qubit-resonator anticrossing can still not be resolved, as the decoherence of the qubits dominates over the coupling.
Nevertheless, the signature of the anticrossing manifests itself in a dispersive shift of the resonance frequency \cite{Oelsner2010,Omelyanchouk2010} and a resulting phase shift.
\begin{figure*}
\includegraphics{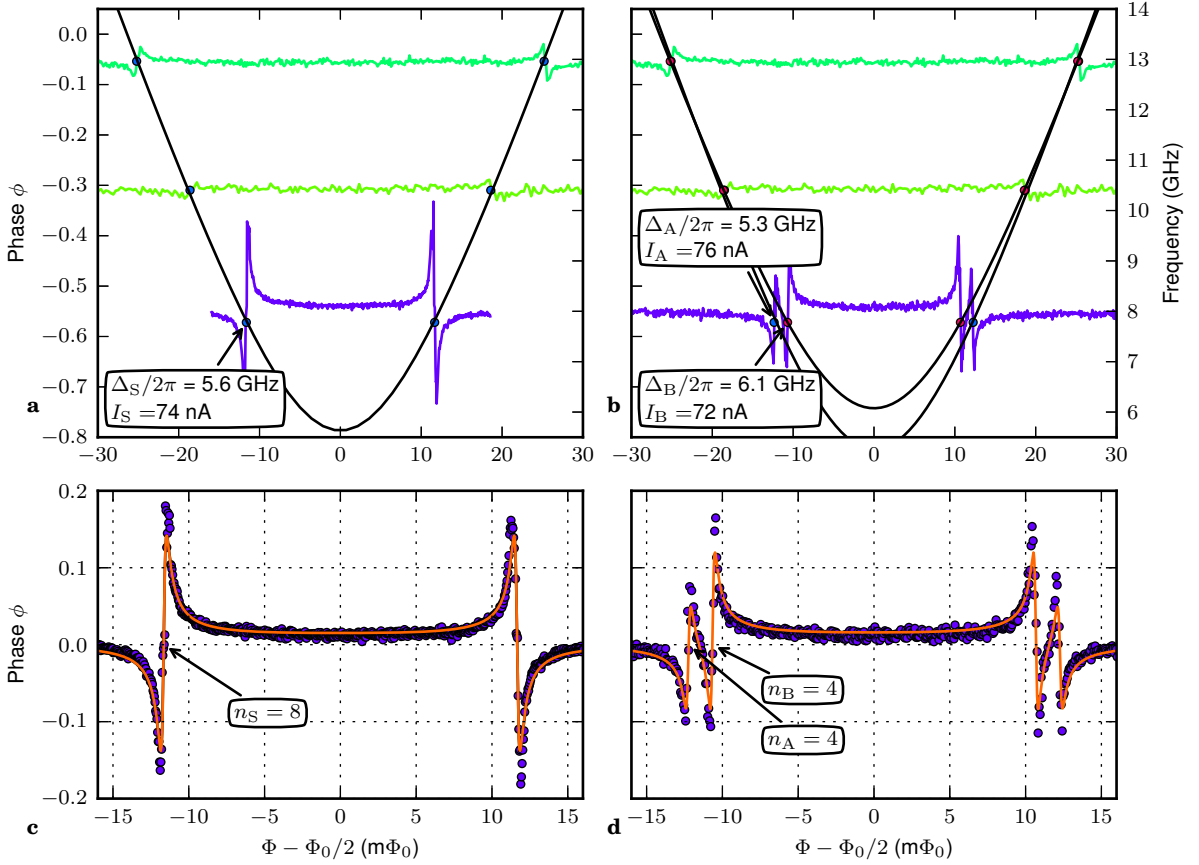}
\caption{\label{fig:main} {\bf The qubit metamaterial in the resonant regime: extraction of ensemble parameters and quantitative analysis of the resonant modes in the third harmonic.} {\bf a,} The phase signal of the transmission through the system at the 3rd, 4th and 5th harmonic in dependence on the magnetic flux, which controls the qubit transition frequencies. The curves are separated in phase by an arbitrary offset. The right y-axis presents the probe frequency of each curve and also corresponds to the qubit frequency (solid line) of the effective parameter set S, which can be mapped at the resonance points.  {\bf b,} The same for the qubit system in the state of two resonant modes. The parameters of ensembles A and B are extracted.  {\bf c,} Quantitative analysis of the resonant mode between the qubits and the third harmonic mode of the resonator. The solid line shows a fit according to Eq. (\ref{eq:FullPhaseShift}) which shows that ensemble S constitutes 8 qubits.  {\bf d,} The best fit for ensembles A and B yields 4 qubits each.}
\end{figure*}
Probing the resonator at frequencies $\omega_1$ and $\omega_2$ 
reveals no resonant interaction between qubits and the resonator (see Supplementary Information B). 
Next, the resonator is probed at its harmonics $\omega_3$, $\omega_4$ and $\omega_5$. 
Two symmetric features, which appear most prominently in the third harmonic signal (see Fig. \ref{fig:main}a), correspond to a resonant interaction between qubits and resonator.
In order to obtain the effective parameters of the qubits belonging to the ensemble that is responsible for the phase shift, the central frequencies of the resonances at harmonics  $\omega_3$ to $\omega_5$ are fitted as a function of magnetic flux to the hyperbolic qubit spectrum (see Fig. 2a). The minimal energy level spacing and persistent current are found to be $\Delta_{\mathrm{S}}/2\pi=5.6$\;GHz and $I_{\mathrm{S}}= (74\pm1)$\;nA. 
Note that these are average values for the individual qubits taking part in the ensemble.
Considering the fourth harmonic at frequency $\omega_4$, the current of the standing wave at the center of the resonator is expected to be zero and the voltage has maximum amplitude. Therefore, the interaction between the qubits and the resonator can only arise from capacitive rather than inductive coupling. We argue that the signature of crossing the qubit spectrum seen at the fourth harmonic is due to the relatively low ratio of Josephson energy to charging energy of the Josephson junctions, leading to non-negligible capacitive coupling between qubits and resonator. As a consequence, the qubits are to be sensitive to charge fluctuations. This may cause low frequency oscillations of the qubit energy leading to the observed splitting of the single resonance into two resonant modes over time, as shown in Fig. \ref{fig:main}b (see also Supplementary Information C). Each of the observed features (Fig. \ref{fig:main}a and Fig. \ref{fig:main}b) are stable over a time scale of $\sim10^3$ s, which is much longer than the typical spectroscopy time of this experiment.  
The effective parameters of the ensembles responsible for the resulting two resonant modes A and B are $\Delta_{\mathrm{A}}/2\pi=5.3$ GHz, $I_{\mathrm{A}}= (76 \pm1 )$ nA and $\Delta_{\mathrm{B}}/2\pi=6.1$ GHz, $I_{\mathrm{B}}= ( 72 \pm 1)$ nA. The coupling of a single qubit to the third harmonic of the resonator is calculated to be $g_{i3}/2\pi = (1.2 \pm 0.1) $ MHz.
If $n$ qubits are in resonance, the stationary phase shift ($\dot{\langle a\rangle}=0$) takes a simple form,
\begin{equation}
\tan{\phi} =  \frac{-2 n {g_{\epsilon,ij}}^2 \delta_{ij}}{\kappa_j	 \left( \Gamma_{\varphi}^2 +  \delta_{ij}^2 \right) + 2n {g_{\epsilon,ij}}^2 \Gamma_{\varphi} }  \;. \label{eq:FullPhaseShift}
\end{equation}
The remaining unknown parameters of the system are the number of qubits $n$ in the ensemble and the dephasing rate $\Gamma_\varphi$. 
The dephasing is responsible for the width of the resonant mode, whereas the dispersive shift out of resonance is independent on the dephasing while depending solely on the number of qubits. Thus, $n$ and $\Gamma_{\varphi}$ can be regarded as independent fitting parameters for the central region and for the periphery of the avoided level crossing, respectively. Still, for small number of qubits $n$, the magnitude of the resonator phase shift (\ref{eq:FullPhaseShift}) depends linearly on $n$ .

The best fit for the most prominent resonant mode (Fig.\;\ref{fig:main}a) between the qubit metamaterial and the third harmonic mode of the resonator yields $n_{\mathrm{S}} = 8$ and $\Gamma_{\varphi,\mathrm{S}} = 2 \pi \cdot 53$\;MHz, as shown by the solid line in Fig.\;\ref{fig:main}c. This dephasing rate corresponds to a phase coherence time of a few ns, as expected for flux qubits operated away from their degeneracy point.
Since the two separated resonant modes (Fig.\;\ref{fig:main}b) are detuned from each other, they can be treated independently. Accordingly, the total phase shift of the signal transmitted through the resonator results from the individual phase shifts induced by ensembles A and B. The best fit of the measured data returns $n_{\mathrm{A}} = 4$ and $\Gamma_{\varphi,\mathrm{A}} = 2 \pi \cdot 54$\;MHz, and $n_{\mathrm{B}} = 4$ and $\Gamma_{\varphi,\mathrm{B}} = 2 \pi \cdot 41$\;MHz (see solid line in Fig. \ref{fig:main}d). The resonant mode of ensemble B is closer to its degeneracy point, which is consistent with a slightly lower dephasing rate. For each of the resonant modes, the number of participating qubits is half of that found for the single resonant mode. Therefore, we conclude that the ensemble of the single resonant mode (ensemble S, Fig.\;\ref{fig:main}a,c) is formed by the same qubits as ensembles A and B (Fig.\;\ref{fig:main}b,d). The remaining qubits, which are not observed as resonant interactions, are still present and contribute to the collective dispersive phase shift in the first two harmonics of the resonator (see Supplementary Information B).

In conclusion, we have reported experiments and analysis of a prototype quantum metamaterial formed by 20 superconducting flux qubits.  The studied example constitutes the first implementation of a basic quantum metamaterial in the sense, that many artificial atoms are collectively coupled to the quantized mode of a photon field.
In spite of the expected relatively large spread of the qubits parameters given by the exponential dependence on the energy gap $\Delta$ and the persistent current $I$ of the qubits, the collective behavior of qubits is clearly observed.
The parameters of three different ensembles of qubits are reconstructed by using their level crossing with the higher harmonics of the resonator. The quantitative analysis of the resonant modes reveals that two ensembles are formed by the collective interaction of 4 qubits each and the third ensemble is formed by 8 qubits. Interestingly, the system exhibits a time dependence, where the large ensemble dissociates into the two smaller ones.

\vspace{1cm}
{\textbf{ Acknowledgements}}
We acknowledge the financial support of the EU through the project SOLID and from the European Community's Seventh Framework Programme (FP7/2007-2013) under Grant No. 270843 (iQIT).PM acknowledges fruitful discussions with Susanne Butz, Pavel Bushev and Boris Ivanov.

\section{Supplementary Information}\label{supp}

\subsection{\bf{A:} Quasiclassical equations of motion}

In this section we will derive the equation of motion (1) for the radiation field 
of the oscillator mode. For the derivation we will also allow qubit-qubit coupling, and 
show that as long as it is smaller than the dephasing rate $\Gamma_{\varphi}$ it is of no relevance. 

The total Hamiltonian of the system is given by,
$ H_T=H+H_{qq}$, where $H$ is the Tavis-Cummings Hamiltonian, shown above, and
 $H_{qq}$ is the qubit-qubit coupling of the form,
\begin{equation}
 H_{qq}=\hbar g_{qq}\sum_i^n(\sigma_+^i\sigma_-^{i+1}+\sigma_{-}^i\sigma_+^{i+1}) \nonumber
\end{equation}
We consider here nearest neighbour coupling, with a coupling strength $g_{qq}$.
This gives us the following equations of motion,
\begin{eqnarray}
 \frac{d}{dt}\langle a \rangle &=&-\frac{\kappa_j}{2}\langle a \rangle - \mathrm{i} g \sum_i^n\langle \sigma_-^i- \rangle + \mathrm{i} \frac{f}{2} \, ,  \nonumber \\
  \frac{d}{dt}\langle \sigma_-^i \rangle &=& - \left(\Gamma_{\varphi} 
  + \mathrm{i} \delta_{ij}\right)\langle \sigma_+^i \rangle + \mathrm{i} g \langle \sigma_z^i a\rangle\nonumber\\ 
   & &  +\mathrm{i} g_{qq}(\langle \sigma_-^{i-1}  \sigma_{z}^i \rangle + \langle \sigma_-^{i+1} \sigma_{z}^i  \rangle )\, ,\nonumber\\
  \frac{d}{dt} \langle \sigma_z^i\rangle &=& -2 \mathrm{i} g \left( \langle \sigma_+^i a\rangle -\langle \sigma_-^i a^{\dag}\rangle\right) 
- \Gamma_1 \left( \langle \sigma_z^i\rangle + 1\right)\nonumber\\
    & &-2\mathrm{i}g_{qq}(\langle \sigma_+^i \sigma_-^{i-1} \rangle  
    -\langle \sigma_-^i \sigma_+^{i-1} \rangle ) \nonumber \\
    & &-2\mathrm{i}g_{qq}(\langle \sigma_+^i \sigma_-^{i+1} \rangle - \langle \sigma_-^i \sigma_+^{i+1} \rangle )\nonumber\\ \nonumber
\end{eqnarray}
With the qubit decay rate $\Gamma_1$ and the dephasing rate $\Gamma_{\varphi}=\Gamma_1/2+\Gamma_{\varphi}^*$, where $\Gamma_{\varphi}^*$
is the pure dephasing rate. 
We seek the solution of the equations of motion in the stationary limit,
 $\dot{\langle a\rangle}=\dot{\langle \sigma_-^i\rangle}=\dot{ \langle \sigma_z^i\rangle}=0$,
and the semiclassical approximation
$ \langle \sigma_k^i a\rangle= \langle \sigma_k^i\rangle \langle a\rangle $.
In the zeroth order of $g_{qq}$ we get,
\begin{eqnarray}
 \langle \sigma_{z}^i\rangle &=& -1\left/\left[1+\frac{4 g^2}{\Gamma_1}\frac{\Gamma_{\varphi}|\langle a\rangle|^2}{\Gamma_{\varphi}^2+\delta_{ij}^2}\right]\right. \nonumber \\
 \langle \sigma_{-}^i\rangle &=& \frac{\mathrm{i} g \langle \sigma_z^i \rangle \langle a \rangle }{\Gamma_{\varphi}+\mathrm{i}\delta_{ij}} \nonumber \;.
\end{eqnarray}
Since the driving is very weak and $|\langle a \rangle|^2 \ll 1$, we can neglect all terms of that order. This directly leads to the equation of motion
which we use to analyse the experiment (1). If we now try to understand the effect of the coupling terms we
see that in the semiclassical approximation we have
\begin{equation}
 \langle \sigma_+^i \sigma_-^{i+1} \rangle=\langle \sigma_+^i \rangle \langle \sigma_-^{i+1} \rangle \propto |\langle a \rangle|^2 \;,\nonumber
\end{equation}
 and these terms can be neglected. Therefore even with qubit-qubit coupling we can assume $\langle \sigma_z\rangle=-1$.
Using this assumption we get the equation of motion,
\begin{eqnarray}
 \frac{d}{dt}\langle \sigma_-^i \rangle &=& - \left(\Gamma_{\varphi} 
  + \mathrm{i} \delta_{ij}\right)\langle \sigma_-^i \rangle - \mathrm{i} g \langle a\rangle\nonumber\\ 
   & &  -\mathrm{i} g_{qq}(\langle \sigma_-^{i-1} \rangle + \langle \sigma_-^{i+1}   \rangle )\, . \nonumber
\end{eqnarray}
From this we get in first order of $g_{qq}$,
\begin{equation}
 \langle \sigma_{-}^i\rangle = \frac{-\mathrm{i} g  \langle a \rangle }{\Gamma_{\varphi}+\mathrm{i}\delta_{ij}}  
                            - \frac{2 g_{qq}\, g   \langle a \rangle }{\left(\Gamma_{\varphi}+\mathrm{i}\delta_{ij}\right)^2} + O(g_{qq}^2)\nonumber      
\end{equation}
Since $\Gamma_{\varphi}\gg g_{qq}$, we see that the effect of qubit-qubit coupling can be neglected.

\subsection{\bf{B:} Phase shift in the dispersive regime}\label{supp:Dispersive}

\begin{figure}
\includegraphics{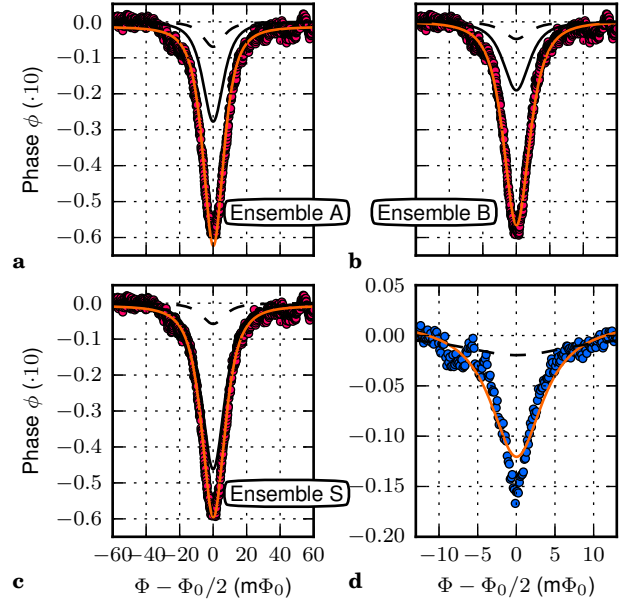}
\caption{\label{fig:MM_dip} {\bf Dispersive shift induced by all qubits at the first two harmonic frequencies in dependence on the magnetic flux}. In each panel, the black dashed line shows the expected curve for a single qubit. {\bf a-c}, the dispersive shift at the fundamental mode frequency. The black solid lines are the theoretical phase shifts for ensembles A, B and S, where the number of qubits $n$ is known. No ensemble accounts for the full shift by itself. The orange solid lines show fits to Eq. (\ref{eq:phase-dispshift}) with $n$ as only fitting parameter. {\bf d}, the dispersive shift at the second harmonic mode frequency. The orange solid line shows a fit to Eq. (\ref{eq:phase-dispshift}) using ensemble parameters S with $g_{i2}$ as only fitting parameter.}
\end{figure}

When all the qubit gaps $\Delta_i$ are higher than the resonator frequency no resonant interaction will occur. A frequency shift - the so-called dispersive dip - of the resonator is observed while tuning the magnetic field. It can be understood as a consequence of the AC-Zeeman shift, where each qubit shifts the cavity frequency by $ g_{\epsilon,ij}^2/\delta_{ij}$ in a positive or negative direction, depending on its state. If the system remains in the ground state at all times, the cavity shift depends solely on the qubit-resonator detuning.

For frequencies below $5.3$ GHz, the qubit metamaterial is in the full dispersive regime, which is the case when probing the resonator at the fundamental mode frequency $\omega_{1}/2\pi$ and at the second harmonic frequency $\omega_{2}/2\pi$.
In this regime the detuning between qubits and resonator $\delta_{ij}$ is always higher than the dephasing $\Gamma_{\varphi}$. The formula for the transmitted phase through the metamaterial consisting of $n$ atoms, Eq. (2), consequently simplifies to
\begin{equation}
\tan{\phi} = - \frac{2 n {g_{\epsilon,ij}}^2}{\kappa_j \delta_{ij}} \;. \tag{S1}
\label{eq:phase-dispshift}
\end{equation}

Fig. \ref{fig:MM_dip}a-c shows the dispersive shift measured at the fundamental mode frequency $\omega_{1}/2\pi$. The experimental data is compared to the expected shifts for the different qubit ensembles (A,B, and S) discussed in the main text. Note, that the shift induced by a single qubit (black dashed lines) is always much weaker than the one actually observed. Parameter sets A and B constitute 4 qubits each. The black solid line shows the theoretical dispersive shift for each ensemble independently (see Fig. \ref{fig:MM_dip}a,b). As expected, they do not account for the full magnitude of the shift. The same is true for set S with 8 qubits in total (see Fig. \ref{fig:MM_dip}c). When fitted to Eq. (\ref{eq:phase-dispshift}) with the ensemble parameters A, B, S (orange solid lines) and $n$ as a free parameter, best fits are obtained for 9, 12 and 10 qubits, respectively. The fit to the parameters for set A slightly overshoots while the one for set B is slightly smaller. In contrast, the fit using the parameters of set S agrees well to the data, which indicates that those parameters reflect the average parameters of the qubit system well. In principle when measuring only the dispersive shift, the state of the system is unknown.

The dominating influence on the dispersive shift arises from the qubits in the resonant modes, which possess a minimal detuning relative to the fundamental mode. The remaining qubits can have a high energy gap (low $\alpha$) resulting in a very small contribution to the dispersive shift which is proportional to $1/{E}^3$. Another explanation for their weak influence may be a very low persistent current or a very small gap, both resulting in a small coupling and therefore a small contribution to the dispersive shift.
However, the full extend of the dispersive shift is induced by all qubits in the metamaterial.

In the second harmonic, the standing wave in the resonator possesses a minimum in the current and a maximum in the voltage at the position of the qubits. The coupling of qubits to the resonator is governed by the capacitance between both. The coupling constant is unknown, but can be determined experimentally from the dispersive shift measured in $\omega_{2}$ (see Fig. \ref{fig:MM_dip}d).  Considering the fit with the coupling as free parameter and the mean values from sets S as well as an effective qubit number of $n = 10$,  a coupling of $g_{i2}/2\pi \approx 0.4$ MHz is obtained. The fit deviates from the data, as it appears to be somewhat steeper and deeper. Ensemble A with a minimal splitting $\Delta_{\mathrm{A}}/2\pi = 5.3$ GHz is fairly close to the second harmonic frequency $\omega_{2}/2\pi = 5.202$ GHz. Although still in the dispersive regime, those qubits lie close to the resonant regime, because the detuning between qubits and resonator is of the same order as the dephasing. This could be the reason for the observed deviation.
	
\subsection{\bf{C:} Time dependence and stability of the system}\label{supp:Time}

The system exhibits two stable states. Below, the dependence on time is described and the stability over time of those two states is shown.

\begin{figure}
\includegraphics[]{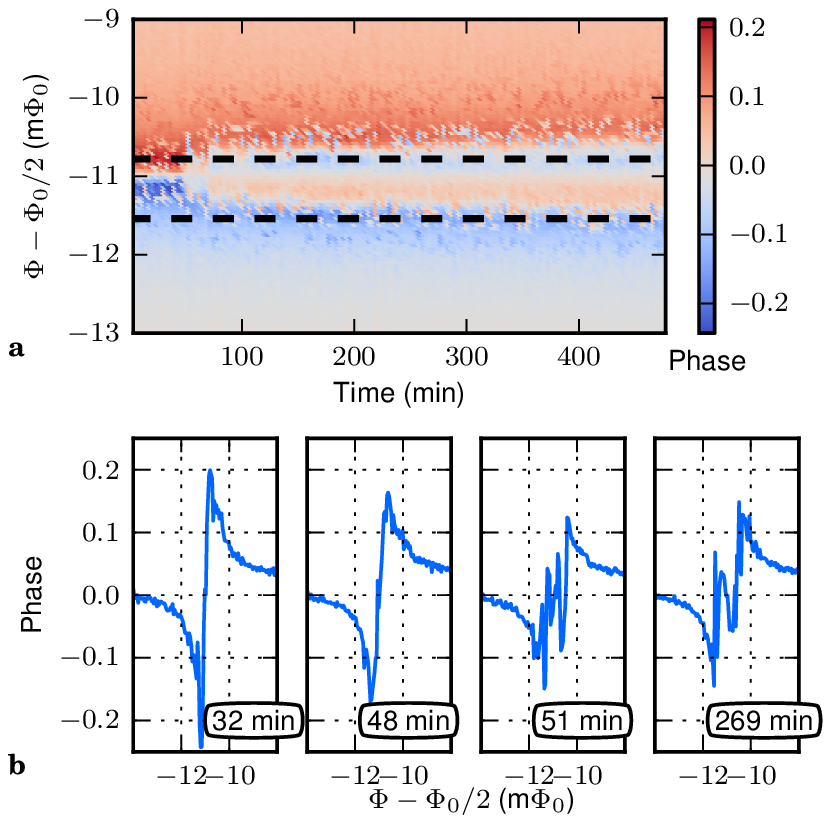}
\caption{\label{fig:suppl_time} {\bf The two states of the system over a time of 7 hours}. {\bf a,} A transition from the state of a single resonant mode to the state of two resonant modes is observed. The black dashed lines are guides to the eyes. {\bf b,} Selected traces from the time dependence. At $t=32$ min and $t=269$ min the two stable states are shown. The two traces in the middle demonstrate the transition from a single resonant mode ($t=48$ min) to two resonant modes ($t=51$ min).}
\end{figure}

Figure \ref{fig:suppl_time}a shows the development of the system over time. The phase at the third harmonic signal frequency is continuously monitored for a fixed flux range. In the beginning of the measurement the system is in the state of a single resonant mode. Each trace is averaged over a period of 3 minutes. The cryostat and the test setup are left undisturbed and no parameters are varied. After about 45 min the transition starts.

First, the magnitude of the resonant mode is reduced. Subsequently, the qubits start to decouple from each other and a state of several resonances is reached. In the end, the system settles in the state of two resonant modes. The full process takes several minutes, such being very short compared to the overall time scale (see Fig. \ref{fig:suppl_time}b). Once the transition is completed the system is again stable over time.
Non-magnetic changes to the gap of the qubits could arrive from its sensitivity to charges, for small ratios of $E_j/E_c$ the gap depends on the voltage across the smallest junction. It is currently under investigation to which degree the gap might change due to charge noise. 

\subsection{\bf{D:} Calculation of the qubit-resonator coupling and its experimental verification}\label{supp:Coupling}

\begin{figure}
\includegraphics[]{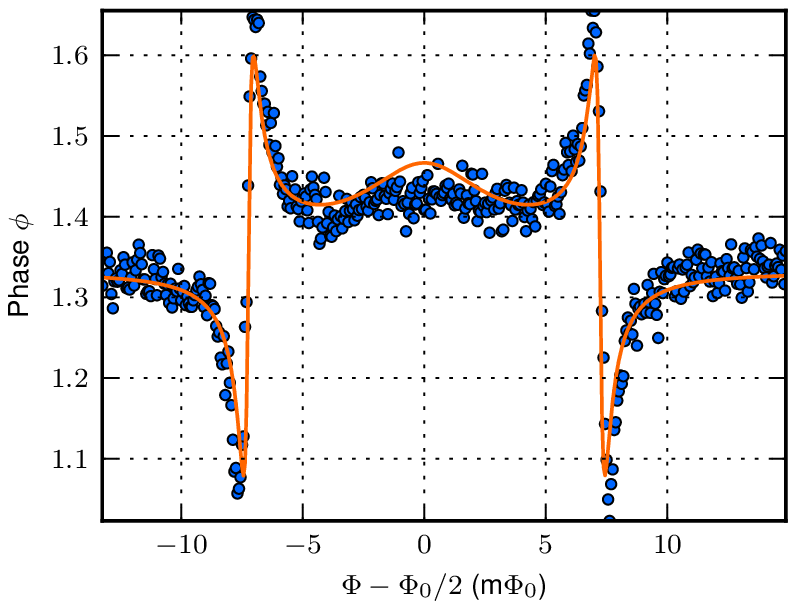}
\caption{\label{fig:suppl_singlequbit} {\bf Transmitted phase through a resonator containing a single flux qubit with known parameters.} 
The solid line shows a two-parameter fit to Eq. (2) for $n=1$. The expected coupling is $g_{qr}/2\pi = (4.7 \pm 0.3)$ MHz, while the best fit is obtained for $\tilde{g}_{qr}/ 2\pi =  4.9$ MHz.}
\end{figure}

The mutual inductance $M_{qr} = 0.51 \pm 0.02$ pH between a single qubit and the resonator and the inductance of the resonator $L_r = (11 \pm 0.4)$ nH have been numerically calculated. The uncertainty for the mutual inductance results from resolution of the micrograph from which the exact position and size of the qubit is extracted.
Our numerical calculation has been experimentally validated using a single qubit embedded into an identical resonator with $\omega_3/2\pi = 7.77$ GHz and $\kappa_3=0.46$ Mhz. The dimension and location of the qubit differs from the ones used in the metamaterial,
the size of the qubit is $l=4.6\;\mu$m and $h=2.6\;\mu$m. It is placed at a distance $x=1.8\;\mu$m,
which results in a slightly higher mutual inductance $M_{qr} = (0.91 \pm 0.02)$ pH. The gap and the persistent current are $\Delta = 3$ GHz and $I = (158 \pm 1)$ nA, determined in a two-tone spectroscopy experiment. The expected coupling is then $g_{qr}/ 2\pi = (4.7 \pm 0.3)$ MHz. Fig.\;\ref{fig:suppl_singlequbit} shows the transmission through the third harmonic of the resonator. As in the case for the metamaterial two symmetric resonance points occur. The solid line shows a two-parameter fit with $g_{qr}$ and $\Gamma_{\varphi}$ as free parameters using Eq. (2) for $n=1$. The best fit is obtained for $\tilde{g}_{qr}/ 2\pi = 4.9$ MHz and $\Gamma_{\varphi} = 2\pi\cdot 141$ MHz. The experimental and theoretical value for the bare coupling are in very good agreement. The higher dephasing compared to the qubits in the metamaterial results from the larger detuning of the flux qubit from its degeneracy point.

\end{document}